\documentclass[aps, pre, reprint,showpacs,groupedaddress,amsmath,amssymb]{revtex4}

\usepackage[dvips]{graphicx}
\usepackage{bm}
\usepackage{textcomp}

\begin{document}


\title{Chemo-Sensitive Running Droplet}


\author{Yutaka Sumino}
\affiliation{Department of Physics, Graduate School of Science, Kyoto University, Kyoto 606-8502, Japan}

\author{Masaharu Nagayama}
\affiliation{Graduate School of Natural Science \& Technology, Kanazawa University, Ishikawa 920-1192, Japan}

\author{Hiroyuki Kitahata}
\affiliation{Department of Physics, Graduate School of Science, Kyoto University, Kyoto 606-8502, Japan}

\author{Shin-ichiro M. Nomura}
\affiliation{Department of Organic Materials, Institute of Biomaterials and Bioengineering, Tokyo Medical and Dental University, Tokyo 101-0062, Japan}

\author{Nobuyuki Magome}
\affiliation{Department of Food and Nutrition, Nagoya Bunri College, Aichi 451-0077, Japan}

\author{Yoshihito Mori}
\affiliation{Department of Chemistry, Ochanomizu University, Tokyo 112-8610, Japan}

\author{Kenichi Yoshikawa}
\email[To whom correspondance should be addressed. Tel:+81-75-753-3812. Fax:+81-75-753-3779. Email:]{yoshikaw@scphys.kyoto-u.ac.jp}
\affiliation{Department of Physics, Graduate School of Science, Kyoto University, Kyoto 606-8502, Japan}


\date{\today}

\begin{abstract}
Chemical control of the spontaneous motion of a reactive oil droplet moving on a glass substrate under an aqueous phase is reported. Experimental results show that the self-motion of an oil droplet is confined on an acid-treated glass surface. The transient behavior of oil-droplet motion is also observed with a high-speed video camera. A mathematical model that incorporates the effect of the glass surface charge is built based on the experimental observation of oil-droplet motion. A numerical simulation of this mathematical model reproduced the essential features concerning confinement of oil droplet motion within a certain chemical territory, and also its transient behavior. Our results may shed light on physical aspects of reactive spreading and a chemotaxis in living things.

\end{abstract}


\maketitle


\section{Introduction}
The spontaneous motion of reactive droplets has attracted considerable attention in relation to energy transduction by living organisms, i.e. chemomechanical energy transduction. In 1978, the motion of a droplet driven by a surface-tension gradient was predicted by Greenspan to explain some aspects of cell cleaving from a physicochemical viewpoint~\cite{30greenspan}. Various kinds of droplet motion driven by a surface-tension gradient have been reported since, where the surface-tension gradient was given by an initial~\cite{15chaudhury, 21bain} or an external \cite{23ichimura} asymmetry in a surface condition. Spontaneous motion in reactive spreading, or a droplet driven by a self-produced surface-tension gradient, has also been noted in various experimental systems~\cite{19bain, 34schmid, 08dossantos, 22bico, 20lee, 28wasan, 17daniel, 32bain, 03kitahta}; for example, a droplet of surfactant on a gold surface~\cite{19bain}, alloy melting~\cite{34schmid}, and a water droplet containing silane on a glass substrate~\cite{08dossantos, 22bico, 20lee}. To interpret these phenomena, extensive theoretical studies on reactive droplets have been performed with the lubrication approximation~\cite{06degennes, 07brochard, 10thiele}.

An oil-water system composed of an organic phase with potassium iodide and iodine and an aqueous phase containing stearyl trimethyl ammonium chloride (STAC) exhibits self-agitation at the oil-water interface, accompanied by spatio-temporal instability of interfacial tension~\cite{11dupeyrat, 33nakache, 14kai, 13kai, 01yoshikawa, 02magome}. It has also been found that the motion of an oil droplet in an oil-water system exhibits chemo-sensitivity~\cite{05nakata}. Based on the measurement of the electrical potential at the oil-water interface in a similar system, the nature of the electrical fluctuation/oscillation has been shown to strongly depend on the chemical properties and the concentration~\cite{yoshikaw1984}. However, the detailed mechanism of this effect of chemical substances has not yet been clarified. Recently, it was proposed that an oil-water system also shows reactive spreading on a glass surface with recovery of the surface condition~\cite{31shioi, 03sumino}.

In the present study, we experimentally observed the chemo-sensitive motion of an oil droplet on a glass substrate that had been partially pretreated with acid. We found that an oil droplet exhibits various behaviors, such as turning-back motion, stopping at a certain position, or slowing down around regions treated with acid. In addition, we also noted that the velocity and shape of the oil droplet show damped oscillation while the droplet moves continuously. However, detailed observations show that the motion of an oil droplet strongly depends on slight changes in the surface condition of the substrate. In terms of the argument given in our previous study~\cite{03sumino}, we expect that the chemo-sensitivity of oil-droplet motion is caused by the condition of the glass surface affected by acid, which inhibits the aggregation of stearyl trimethyl ammonium ion (STA$^{+}$ ion) on the glass surface. To confirm this hypothesis, we idealize an oil droplet as self-propelling spring-beads whose driving force is given by the surface-tension gradient. This model successfully reproduces the observed experimental results. Furthermore, a numerical simulation of the model reveals that the selection among the turning-back motion, stopping at a certain position, and slowing down can be explained in terms of the effect of the acid on the glass substrate, where minute changes in the acid effect drastically change the behavior of the droplet. As has been reported previously, the oil-water system exhibits temporal fluctuation in electrochemical potential, which is sensitive to chemical stimuli~\cite{yoshikaw1984}. On the other hand, the present study shows that it may be possible to acquire spatio-temporal information on minute changes in the chemical concentration on a surface by noting a drastic change in the characteristic motion of a self-running oil droplet. This result may be associated with the mechanism of chemotaxis in biological systems, where living organisms show extreme sensitivity to a small spatial gradient in chemicals~\cite{mato_1975, carole_1999}.

\begin{figure}
\includegraphics[width=6.5cm,height=7.6cm,keepaspectratio,clip]{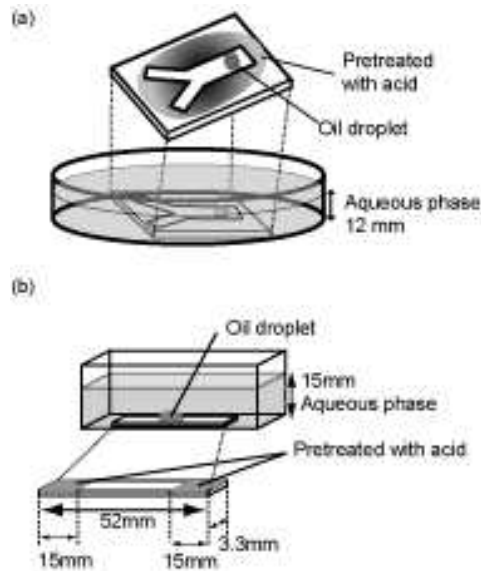}
\caption{Schematic representation of the experimental setup (a) on a Y-shaped path on a glass surface, and (b) on an acid-treated narrow glass substrate. Prior to the experiment, both ends of the narrow glass substrate ($\approx$ 15 mm) were treated with acid.}
\label{fig1}
\end{figure}

\section{Experiments}
The aqueous phase contained 1 mM STAC and the organic phase was 5 mM iodine solution of nitrobenzene saturated with potassium iodide. STAC was prepared by recrystallization using acetone. For glass substrates, we used micro slide glass (Matsunami, Osaka; S9111). The glass surface was treated with acid as follows. First, 1 M sulfuric acid solution was dabbed on the glass substrate with a cotton swab, and then it was rinsed from the surface with running distilled water. Extreme care was taken in rinsing to avoid invading the bare glass surface.

In the measurement on a Y-shaped path, 100 $\mu$l of the oil was placed in an aqueous phase on the pretreated glass substrate [Fig. 1(a)]. A digital video camera (Panasonic; NV-GS100K-K) was used to record the motion of an oil droplet at 30 frames per second.

In the experiments on an acid-treated narrow glass substrate, the glass substrate (1 mm $\times$ 52 mm $\times$ 3.1 - 3.2 mm) was cut from slide glass. Both ends of this substrate were pretreated with acid for 15 mm [Fig.~\ref{fig1}(b)]. An oil droplet of 15-60 $\mu l$ was placed on the substrate in the same manner as in the Y-shaped path experiment. The motion of an oil droplet was recorded by a high-speed video camera (RedLake MASD Inc.,San Diego, CA; Motion Scope PCI) at 125 frames per second. All measurements were carried out at room temperature (20$^{\circ }$C $\pm$3$^{\circ}$C). The recording of a droplet motion was digitized by an image-processing system (Library, Tokyo). 

\begin{figure}
\includegraphics[width=7.5cm,height=11cm,keepaspectratio,clip]{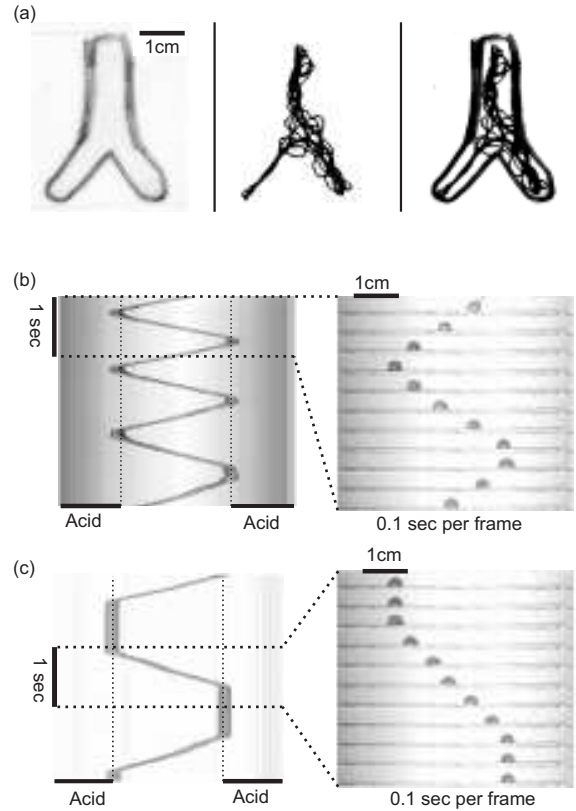}
\caption{Experiments on spontaneous droplet motions. (a) Confined motion of an oil droplet. The path of an oil droplet (100 $\mu$l) was enclosed within the acid-treated surface. The left image shows the actual appearance of a substrate. The outside of the Y-shaped region was treated with acid. In the middle image, the path of the CM of an oil droplet is shown. The superposition of both images is shown on the right. (b),(c) Motion of a 15-$\mu$l oil droplet on a narrow acid-treated glass substrate. An image from the experiment and a spatio-temporal image of oil-droplet motion are shown. An oil droplet showed (b) shuttling motion and (c) intermittent shuttling motion. When an oil droplet did not intrude into the acid-treated region it showed turning-back motion, which resulted in shuttling motion.
}
\label{fig2_1}
\end{figure}

\begin{figure}
\includegraphics[width=7.5cm,height=7.5cm,keepaspectratio,clip]{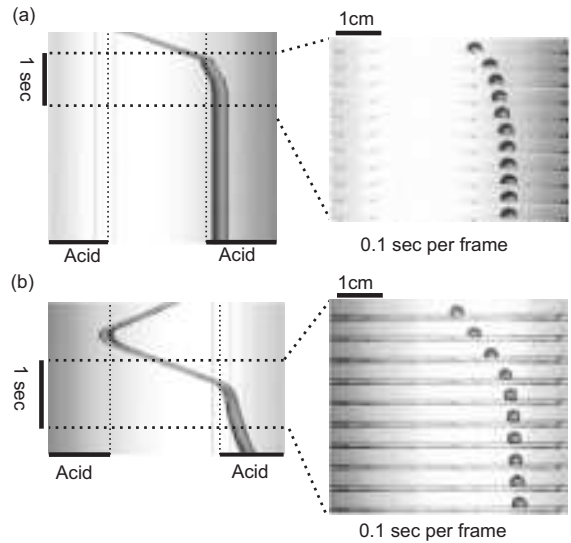}
\caption{Experiments on spontaneous droplet motion. Motion of a 15-$\mu$l oil droplet on a narrow acid-treated glass substrate. An image from the experiment and a spatio-temporal image of oil-droplet motion are shown. An oil droplet showed (a) stopping and (b) slowing. When an oil droplet intruded into the acid-treated region, it showed stopping (a) or slowing (b).
}
\label{fig2_2}
\end{figure}

\section{Results}
\begin{figure}
\includegraphics[width=7.5cm,height=8.5cm,keepaspectratio,clip]{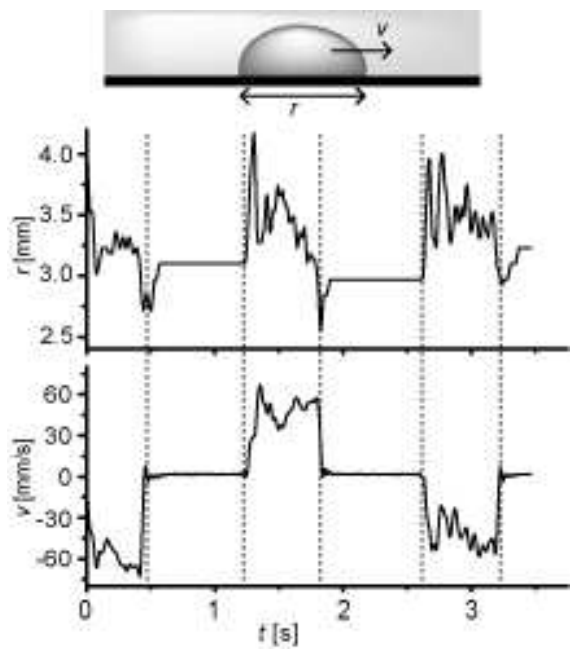}
\caption{Time trace of the apparent length, $r$, and velocity, $v$, of an oil droplet for the motion shown in Fig.~\ref{fig2_1}(c).}
\label{fig3}
\end{figure}

The spontaneous motion of an oil droplet was found to be confined within a Y-shaped path and a narrow substrate [Figs.~\ref{fig2_1},~\ref{fig2_2}]. On the Y-shaped path [Fig.~\ref{fig2_1}(a)], motion is clearly limited to within the region that was not treated with acids. On the narrow substrate, oil droplets exhibit quasi 1-dimensional motion~\cite{03sumino}, i.e. an oil droplet of 15 $\mu$l showed turning-back motion that resulted in a shuttling motion [Fig.~\ref{fig2_1}(b),(c)] when the oil droplet could not intrude into the acid-treated region. The time trace of oil-droplet velocity $v$ and the apparent length of an oil droplet $r$ in the motion shown in Fig.~\ref{fig2_1}(b) are plotted in Fig.~\ref{fig3}. $v$ and $r$ showed damped oscillation after each returning motion until both reached steady values. On the other hand, when an oil droplet of 15 $\mu$l intruded into the acid-treated region, the oil droplet stopped and stood still [Fig.~\ref{fig2_2}(a)] or slowed down [Fig.~\ref{fig2_2}(b)]. The motion of an oil droplet was found to be sensitively dependent on the manner of acid treatment.

\section{Discussion}
\begin{figure}
\includegraphics[width=7cm,height=6cm,keepaspectratio,clip]{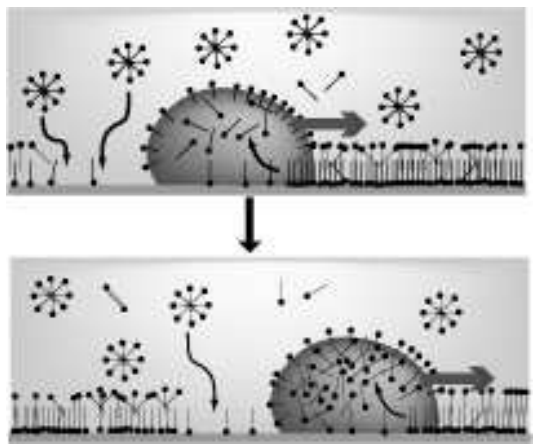}
\caption{Schematic diagram of oil-droplet motion and glass surface conditions. STA$^{+}$ is represented as a hydrophobic bar with a hydrophilic head.}
\label{fig4}
\end{figure}

It has been suggested by our previous study~\cite{03sumino} that the motion of an oil droplet can be explained in terms of reactive spreading~\cite{07brochard}. STA$^{+}$ ions in the aqueous phase aggregate on the glass surface that has negative charges at neutral pH~\cite{atkin03}. The organic phase dissolves STA$^{+}$ ions that have adhered to the glass substrate, and this dissolution is promoted since STA$^{+}$ and I$^{-}_{3}$ in the organic phase make an ion-pair. The ion-pair dissolves well in non-polar solvent, i.e. the organic phase. Therefore, a difference in interfacial energy arises from the difference in the concentration of STA$^{+}$ ions on the glass surface between the front and back of an oil droplet [Fig.~\ref{fig4}], and the oil droplet is driven by Marangoni effect. Thus, the condition of the glass surface on which STA$^{+}$ ions aggregate plays an important role in the motion of an oil droplet.

Nakata and coworkers examined the chemo-sensitivity of the motion of an oil droplet escaping from acids~\cite{05nakata}. In their experiment, hydrochloric acid was injected in the aqueous phase while an oil droplet was moving. From the above discussion, modification of a glass surface by acids significantly affects oil-droplet motion. It has been reported that the point of zero charge (pH$_{0}$) of the SiO$_{2}$ surface is around $2.2$~\cite{01parks, 02yoon}. Thus, the glass surface charge may have been partially reversed to be positive due to the effect of acid since the pH of 1M sulfuric acid is around $0$. We can consider that this results in fewer adsorption sites for STA$^{+}$ ions on the glass surface. This may explain why an oil droplet showed confined motion [Fig.~\ref{fig2_1}(a)] and shuttling motion [Fig.~\ref{fig2_1}(b)] on an acid-treated glass substrate when it was encircled by an acid-treated glass surface. Furthermore, the dependence of droplet motion on the condition of the acid-treated region can be attributed to the difference in the number of adsorption sites on the glass surface. To explain the various motions of an oil droplet, it seems appropriate to control the surface condition of a glass substrate much more accurately. On the other hand, in the experimental process, these different types of the characteristic motions were observed even under nearly identical conditions, including the concentration and duration of treatment with sulfuric acid. To understand this strong dependence on the condition of the glass surface, we constructed a mathematical model, and conducted a numerical simulation of oil-droplet motion on a narrow substrate.

There may be various ways to theoretically analyze for the spontaneous motion of an oil droplet, such as interface dynamics based on Young's equation~\cite{06degennes, 07brochard} or a hydrodynamic approach which adopts the lubrication approximation~\cite{10thiele}. Instead of making a detailed model, we propose a simple ``beads-spring" model for an oil droplet to abstract the essential features of droplet motion.

\section{Modeling}
\begin{figure}
\includegraphics[width=7.5cm,height=3.5cm,keepaspectratio,clip]{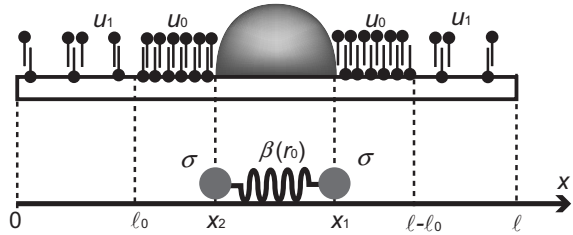}

\caption{Schematic diagram of the model of oil-droplet motion on a glass surface. The lower arrow represents the spatial coordinate, where $0$ to $\ell_{\mathrm{0}}$ and $\ell-\ell_{\mathrm{0}}$ to $\ell$ are treated with acid. $u_{\mathrm{0}}$ and $u_{\mathrm{1}}$ are the saturated concentration of the STA$^{+}$ ions on the glass substrate on bare and acid-treated glass substrate, respectively. The oil droplet is modeled as two lines (perpendicular to the $x$-axis), with a mass density of $\sigma$, connected to a spring with an elastic constant of $\beta(r_{\mathrm{0}})$ (see also Appendix A)}
\label{schematic}
\end{figure}

Since we want to know the basic mechanism of the motion of an oil droplet on a glass substrate, we make a simple one-dimensional model. The $x$ axis is set along the surface of the glass substrate. The droplet proceeds on a line with stretching and shrinking. The positions at the both ends of the droplet are set as $x_1(t)$ and $x_2(t)$, where $x_1 < x_2$. The concentration of STA$^{+}$ ions on the glass surface is set as $u(x,t)$. The dynamics of $x_1$ and $x_2$ are written as follows:
\begin{equation}\label{D02}
\sigma \frac{{\rm d}^2 x_1}{{\rm d} t^2} = - \mu_0 \frac{{\rm d} x_1}{{\rm d} t} - \left. \frac{\partial E}{\partial x} \right|_{x = x_1} - \beta(r_{\mathrm{0}})(x_1(t) - x_2(t) - r_{\mathrm{0}}),
\end{equation}
\begin{equation}\label{D02+}
\sigma \frac{{\rm d}^2 x_2}{{\rm d} t^2} = - \mu_0 \frac{{\rm d} x_2}{{\rm d} t} - \left. \frac{\partial E}{\partial x} \right|_{x = x_2} - \beta(r_{\mathrm{0}})(x_2(t) - x_1(t) + r_{\mathrm{0}}),
\end{equation}
where $\sigma$, $\mu_0$, and $r_{\mathrm{0}}$ correspond to the mass of the oil droplet reduced to two lines, a viscous damping coefficient and the characteristic size of the oil droplet, respectively. $\beta(r_{\mathrm{0}})$ is the elastic constant of the oil droplet with characteristic size $r_{\mathrm{0}}$ (see Appendix A):
\begin{equation}
\beta(r_{\mathrm{0}}) = \frac{2 \gamma w}{r_{\mathrm{0}}}.
\end{equation}
$E(u)$ is the surface energy of the glass surface modulated by the STA$^{+}$ ions adsorbed on it:
\begin{equation}\label{D03}
E(u) = \frac{e_0}{1 + a u^n},
\end{equation}
where $e_0$, $a$ and $n$ are positive constants.
On the other hand, the dynamics of $u$ can be written as follows:
\begin{equation}\label{D04}
\frac{\partial u}{\partial t} = d \frac{\partial^2 u}{\partial x^2}  + F(u, x, x_1(t), x_2(t); \ell_0),
\end{equation}
where $d$ is the diffusion constant of the STA$^{+}$ ions on the glass surface. $F(u, x, x_1(t), x_2(t); \ell_0)$ corresponds to the desorption of STA$^{+}$ ions from the glass surface to the oil droplet. For this term, we assume the following description:
\begin{align}\label{D06}
&F(u, x, x_1(t), x_2(t); L_0) \notag \\
&= \left\{ \begin{array}{ll}
	- k_1 u , & (x_2(t) \le x \le x_1(t)) \\
	k_2(u_0 - u), & (\ell_0 < x < x_2(t), \, x_1(t) < x < \ell - \ell_0) \\
	k_3(u_1 - u), & (0 < x \le \ell_0, \, \ell - \ell_0 \le x < \ell)
\end{array} \right. 
\end{align}
\begin{figure}
\includegraphics[width=7.5cm,height=10.5cm,keepaspectratio,clip]{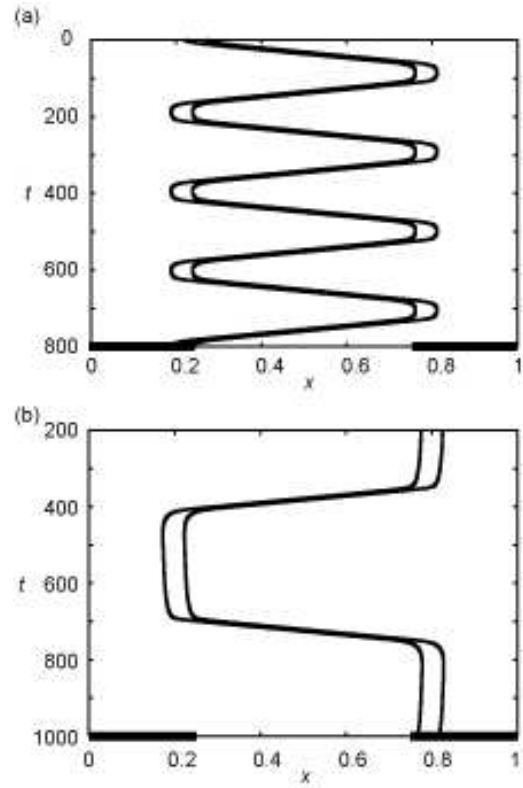}
\caption{Spatio-temporal plot of the oscillatory motion of a droplet given by numerical calculations using Eqs. (\ref{D10}) - (\ref{D19}). Only the parameter $U_1$ was changed: (a) $U_1 = 0.65$ and (b) $U_1 = 0.658$. The other parameters are $\mu = 0.2$, $R_{\mathrm{0}} = 0.05$, $E_0 = 5 \times 10^{-5}$, $A = 1$, $n = 4$, $W = 0.1$, $D = 2.5 \times 10^{-5}$, $L_0 = 0.25$, $U_0 = 1$, $K_2 = 1$, $K_3 = 1$, $X_{10} = 0.23$, and $X_{20} = 0.27$. The thick lines along the $x$ axis correspond to the regions treated with acid.}
\label{fig5}
\end{figure}
where $k_1$, $k_2$, $u_0$, $\ell$ and $\ell_0$ are positive constants that correspond to the desorption rate, the adsorption rate, the saturated concentration of STA$^{+}$ ions on the glass surface, the length of the glass substrate, and the length of the glass substrate treated with acid, respectively (see Fig.~\ref{schematic} and Appendix B). The initial and boundary conditions are set as
\begin{equation} \label{D07}
\frac{\partial u}{\partial x}(t, 0) = \frac{\partial u}{\partial x}(t, \ell) = 0,
\end{equation}
\begin{figure}
\includegraphics[width=7.5cm,height=10.5cm,keepaspectratio,clip]{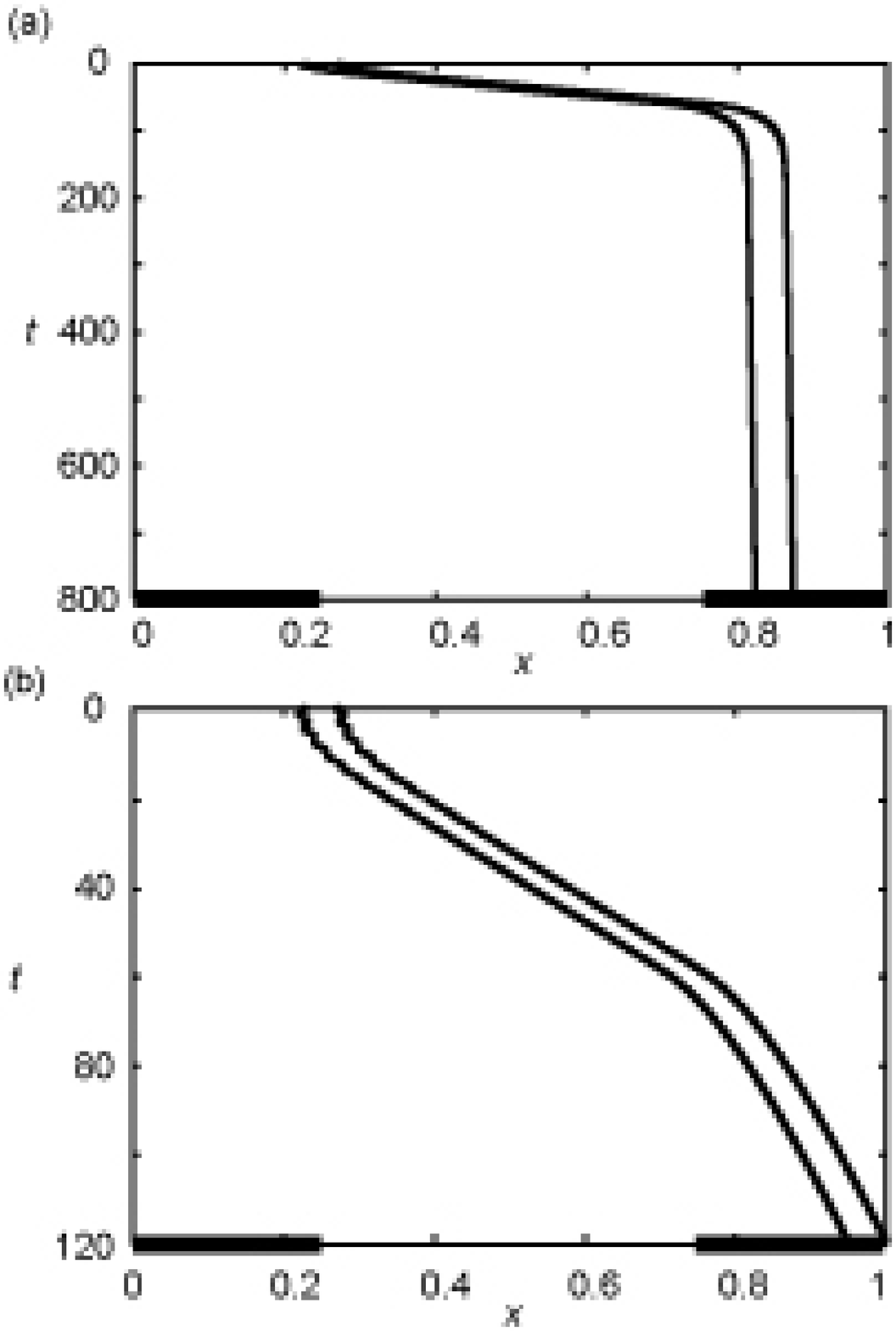}
\caption{Spatio-temporal plot of the slowing of a droplet in the acid-treated region given by numerical calculations using Eqs. (\ref{D10}) - (\ref{D19}). Only the parameter $U_1$ was changed: (a) $U_1 = 0.68$ and (b) $U_1 = 0.75$. The other parameters are the same as those in Fig. \ref{fig5}. The thick lines along the $x$ axis correspond to the regions treated with acid.}
\label{fig6}
\end{figure}
\begin{align} \label{D08}
u(0, x) = \left\{
\begin{array}{ll}
u_0, & (\ell_0 < x < \ell - \ell_0) \\
u_1, & (0 < x \le \ell_0, \, \ell - \ell_0 \le x < \ell)
\end{array} \right.
\end{align}
\begin{align}\label{D09}
x_1(0) = x_{10}, \, \, \, & x_2(0) = x_{20}( < x_{10}), \notag \\
\frac{{\rm d} x_1}{{\rm d} t}(0) &=  \frac{{\rm d} x_2}{{\rm d} t}(0) = 0.
\end{align}

To normalize Eqs. (\ref{D02}) - (\ref{D09}), we introduce the following dimensionless variables:
\begin{equation}
U=\frac{u}{u_0}, \, \, X_1=\frac{x_1}{\ell}, \, \, X_2 = \frac{x_2}{\ell}, \, \, X = \frac{x}{\ell}, \,\, T = k_1 t.
\end{equation}
We then derive the following dimensionless system from Eqs. (\ref{D02}) - (\ref{D09}):
\begin{align}\label{D10}
\frac{{\rm d}^2 X_1}{{\rm d} T^2} =& - \mu \frac{{\rm d} X_1}{{\rm d} T} - \left. \frac{\partial E}{\partial X} \right|_{X = X_1} \notag \\
 &- B(R_{\mathrm{0}})(X_1(T) - X_2(T) - R_{\mathrm{0}}),
\end{align}
\begin{align}
\frac{{\rm d}^2 X_2}{{\rm d} T^2} =& - \mu \frac{{\rm d} X_2}{{\rm d} T} - \left. \frac{\partial E}{\partial X} \right|_{X = X_2} \notag \\
 &- B(R_{\mathrm{0}})(X_2(T) - X_1(T) + R_{\mathrm{0}}),
\end{align}
\begin{equation}\label{D11}
E(U) = \frac{E_0}{1 + A U^n},
\end{equation}
\begin{equation}\label{D12}
\frac{\partial U}{\partial T} = D \frac{\partial^2 U}{\partial X^2}  + F(U, X, X_1(T), X_2(T); L_0),
\end{equation}
\begin{align}
\label{D13}
&F(U, X, X_1(T), X_2(T); L_0) \notag \\
&= \left\{ \begin{array}{ll}
	- U , &  (X_2(T) \le X \le X_1(T)) \\
	K_2(U_0 - U), & (L_0 < X < X_2(T), \\
	& \,\,\, X_1(T) < X < 1 - L_0) \\
	K_3(U_1 - U), & (0 < X \le L_0, \, 1 - L_0 \le X < 1)
\end{array} \right. 
\end{align}
with the following initial and boundary conditions:
\begin{equation} \label{D17}
\frac{\partial U}{\partial X}(T, 0) = \frac{\partial U}{\partial X}(T, 1) = 0,
\end{equation}
\begin{align} \label{D18}
U(0, X) = \left\{
\begin{array}{ll}
1, & (L_0 < X < 1 - L_0) \\
U_1, & (0 < X \le L_0, \, 1 - L_0 \le X < 1)
\end{array} \right.
\end{align}
\begin{align}\label{D19}
X_1(0) = X_{10}, \, \, \, & X_2(0) = X_{20}( < X_{10}), \notag \\ \frac{{\rm d} X_1}{{\rm d} T}(0) &= \frac{{\rm d} X_2}{{\rm d} T}(0) = 0,
\end{align}
where
\begin{displaymath}
\mu = \frac{\mu_0}{\sigma k_1}, \, \, \, E_0 = \frac{e_0}{\ell \sigma k_1^2}, \,\, R_{\mathrm{0}} = \frac{r_{\mathrm{0}}}{\ell},
\end{displaymath}
\begin{displaymath}
B(R_{\mathrm{0}}) = \frac{\ell^2 \beta(r_{\mathrm{0}})}{\sigma k_1^2} = \frac{W}{R_{\mathrm{0}}}, \, \, W = \frac{2 \gamma w \ell}{\sigma k_1^2 }, \, \, A = a u_0^n,
\end{displaymath}
\begin{displaymath}
D = \frac{d}{k_1 \ell^2}, \, \, K_2 = \frac{k_2}{k_1}, \, \, K_3 = \frac{k_3}{k_1}, \,\, U_1 = \frac{u_1}{u_0},
\end{displaymath}
\begin{displaymath}
L_0 = \frac{\ell_0}{\ell} ( < 1), \, \, X_{10} = \frac{x_{10}}{\ell}, \, \, X_{20} = \frac{x_{20}}{\ell}.
\end{displaymath}

\section{Numerical Results}

\begin{figure}
\includegraphics[width=7.5cm,height=10.5cm,keepaspectratio,clip]{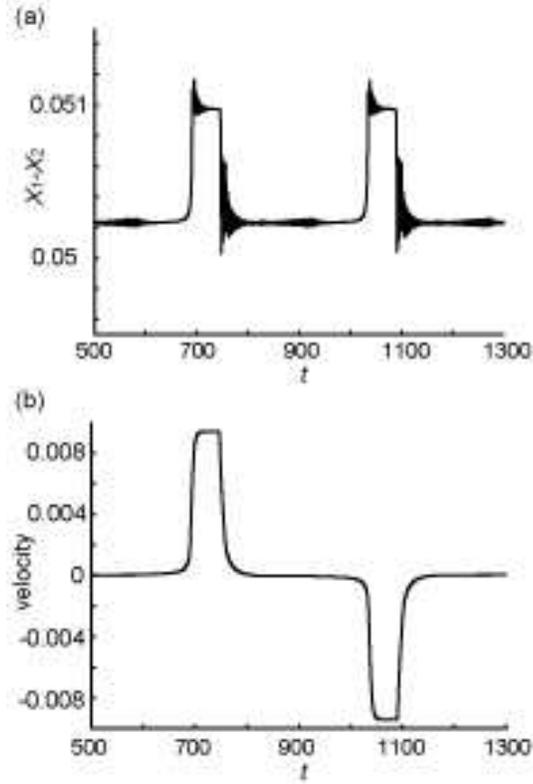}
\caption{Temporal changes in (a) the length $X_{1}-X_{2}$ and (b) the velocity of the oil droplet undergoing the oscillatory motion shown in Fig. \ref{fig5}(b). The high-frequency oscillation is attributed to the eigen-frequency of the mathematical model that did not appear in the experimental results.}
\label{fig7}
\end{figure}

We performed a numerical simulation using Eqs. (\ref{D10}) - (\ref{D13}) under the initial and boundary conditions shown in Eqs. (\ref{D17}) - (\ref{D19}).

We only changed the parameter $U_1$, while all of the other parameters are fixed. It is noted that $U_1$ corresponds to the saturated concentration at the region treated by acid. Spatio-temporal plots of the droplet given by the numerical calculations are shown in Figs.~\ref{fig5} and~\ref{fig6}. 

When $U_1$ is small enough, i.e., it is difficult for STA$^{+}$ ions to adsorb to the acid-treated region, the oil droplet moves back and forth inside the untreated region, as shown in Fig.~\ref{fig5}(a). As $U_1$ increases slightly, the oil droplet does not turn back quickly but rather rests at the border between the acid-treated region and the untreated region. After a brief stop, it begins to move backward, and this intermittent shuttling motion is repeated as shown in Fig.~\ref{fig5}(b). For much larger $U_1$, the oil droplet does not go back any more, as shown in Fig.~\ref{fig6}. The oil droplet goes into the acid-treated region and it continues to move very slowly or stops according to the value of $U_1$.

These differences in the characteristics of the motion of an oil droplet are seen only with slight differences in $U_1$. Thus, the numerical results suggest that a slight fluctuation in acid treatment can cause a dramatic change in the characteristics of the motion of an oil droplet. In the experiments, the oscillatory motion shown in Fig.~\ref{fig2_1}(b), the stopping of the droplet at the border shown in Fig.~\ref{fig2_2}(a), and the slow motion in the acid-treated region shown in Fig.~\ref{fig2_2}(b) were observed under almost the same conditions. We can guess that the various kinds of motion observed in the experiments are due to slight differences in the acid treatment of the glass.

We have also noted that intermittent-oscillatory motion (Fig.~\ref{fig5}(b)) and stopping (Fig.~\ref{fig6}(a)) does not appear without a diffusion term in Eq.~(\ref{D04}). This suggests that the diffusion on a glass surface has a significant effect in the real experimental system. There have been several studies concerning the effect of diffusion on the motion of a self-running droplet~\cite{09shanahan, 25shanahan, 10thiele}. Thiele et al.~\cite{10thiele} stated that in their model a running droplet can be stopped by increasing the effect of a surfactant diffusion on a substrate. In our study, the diffusion of surfactant on the glass substrate is not large enough to stop the motion of a droplet. However, at the boundary of the acid-treated region where an oil droplet stopped, the concentration of the surfactant on the acid-treated region is in equilibrium with the diffused surfactant from the non-treated region. Thus, the effect of diffusion is optimized by the inhomogeneity in the glass surface so that an oil droplet stops its motion.

Figure \ref{fig7} shows (a) the normalized length and (b) the velocity of an oil droplet. When the velocity of the oil droplet is large, the length of the droplet increases, and vice versa, which well corresponds to the actual experimental trend (Fig.~\ref{fig3}).

\section{Conclusion}

In the present study, we investigated the self-running motion of an oil droplet on an acid-treated glass substrate. In these experiments, the oil droplet undergoes either oscillatory motion within the untreated region or stopping/slowing at the border between the acid-treated region and the untreated region. We made a simple ``beads-spring" model and performed numerical calculations, which suggest that a slight difference in the nature of adsorption in the acid-treated region can cause a dramatic change in the characteristics of this motion. The numerical results well reproduced the experimental results. In particular, the numerical results suggest that a slight difference in acid treatment can cause a dramatic change in the motion of an oil droplet. The results with the model also suggest that diffusion on the glass surface plays an important role in the spontaneous motion of a droplet. The inhomogeneity of the glass surface, i.e. the boundary of the surface condition, optimizes the effect of diffusion.

In this study, an oil droplet avoided the area that was treated with acid. It is possible that some other chemicals may attract an oil droplet. Above all, the present results showed that a running droplet senses information regarding a minute variation in the chemical concentration in space by showing a drastic change in its motion. Thus, this result may inspire studies on chemotaxis and chemosensitivity in chemistry as well as biology.

\begin{acknowledgments}
We thank Professor Satoshi Nakata (Nara University of Education) and Doctor Takatoshi Ichino (Kinki University) for their helpful discussions on this work.
This work was supported in part by Grants-in-Aid for the 21st Century COE ``Center for Diversity and Universality in Physics" and for Young Scientists (B) to M. N. from the Ministry of Education, Culture, Sports, Science and Technology of Japan.
\end{acknowledgments}

\appendix

\section{Validity of the beads-spring model}

\begin{figure}
\includegraphics[width=7cm,height=6cm,keepaspectratio,clip]{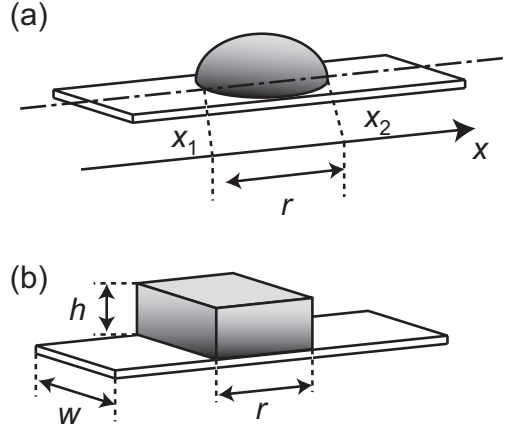}
\caption{Schematic representation of the framework of the ``beads-spring" model. The hemisphere-shaped oil droplet as shown in (a) is regarded as a rectangular solid. Since we consider only the $x$ dimension, the conservation law for the volume corresponds to the elastic constant of the spring.}
\label{figa}
\end{figure}

We discuss here the validity of the beads-spring model for an oil droplet. As shown in Fig.~\ref{figa} , we assume that the droplet is shaped as a rectangular solid, whose length, width and height are $r$, $w$, and $h$, respectively. The volume of the droplet is conserved:
\begin{equation}
V = rwh.
\end{equation}

The main terms of the energy of the droplet are due to gravity and interfacial tension:
\begin{align}\label{AA02}
F =& \frac{1}{2}mgh + \gamma[ r w + 2 h w +2 r h ] \notag \\
= & - \frac{\rho V^2 g}{2 r w} + \gamma[ r w + 2 V ( \frac{1}{r} + \frac{1}{w})],
\end{align}
where $\gamma$ is the interfacial tension, $\rho$ is the density, $m$ is the mass of the droplet, and $g$ is the gravitational acceleration. 

First, we calculate the fixed point. At the fixed point, ${\rm d}F/{\rm d}r = 0$; therefore,
\begin{equation}
\frac{{\rm d}F}{{\rm d}r} = -\frac{\rho V^2 g}{2 r^2 w} + \gamma[ w - \frac{2 V}{r^2} ]= 0,
\end{equation}
which gives the fixed point, $r_0$:
\begin{equation}
r_{\mathrm{0}} = \left(\frac{\rho V^2 g}{2 \gamma w^2} + \frac{2V}{w} \right)^{1/2}.
\end{equation}
The elastic constant, $\beta(r_{\mathrm{0}})$ is given as
\begin{equation}
\beta(r_{\mathrm{0}}) = \left. \frac{{\rm d}^2 F}{{\rm d} r^2} \right|_{r = r_0} = \frac{2 \gamma w}{r_{\mathrm{0}}}.
\end{equation}

From the above discussion, we can regard the oil droplet as two beads connected with a spring whose elastic constant is proportional to the inverse of the characteristic size of the droplet.

The spring constant is on the same order of magnitude even if the shape of the droplet is regarded as a cap-like or ribbon-shaped droplet.

\section{Model of the adsorption/desorption process on a glass surface.}
The effect of acids is thought to result in a decrease in the number of adsorption sites $\mathrm{X^{-}_{sur}}$ for STA$^{+}$ ions ~\cite{atkin03}. Assuming that adsorption-desorption is a langmuir isothermal process,
\begin{equation}
\label{S01}
\mathrm{STA^{+}}+\mathrm{X^{-}_{sur}} \rightleftharpoons \mathrm{STAX_{sur}}.
\end{equation}
Since $m_0$ is the total density of adsorption sites on the glass surface, $x$ is the density of free adsorption sites, and $u$ is the concentration of STA$^{+}$ ion in the aqueous phase, the surface concentration $u_{\mathrm{sur}}$ of $\mathrm{STAX_{sur}}$ under aqueous phase changes according to
\begin{equation}
\label{S02}
\frac{\mathrm{d} u_{\mathrm{sur}}}{\mathrm{d} t} = k u \{x_{0}-(1+\frac{k^{-1}}{k u}) u_{\mathrm{sur}}\},
\end{equation}
where we assume that $x+u_{\mathrm{sur}}=x_{0}$, and $k$ and $k^{-1}$ are reaction constants in Eq. (\ref{S01}).
The effect of acid is included in the decrease in $x_{0}$, which is equivalent to the decrease in $u_0$ to $u_1$ in the acid-treated region in Eq.~(\ref{D06}).

Similarly, the desorption that occurs under the organic phase is considered as, 
\begin{equation}
\label{S03}
\mathrm{STAX_{sur}} + \mathrm{I^{-}_{3}} \rightarrow \mathrm{STAI_{3}} + \mathrm{X^{-}_{sur}}.
\end{equation}
We assume that the reaction proceeds only from left to right.
Therefore, $u_{\mathrm{sur}}$ changes according to
\begin{equation}
\label{S04}
\frac{\mathrm{d} u_{\mathrm{sur}}}{\mathrm{d} t} = -\hat{k} u_{\mathrm{sur}} v,
\end{equation}
where $v$ represents the concentration of I$^{-}_{3}$ in an oil droplet, and $\hat{k}$ represents the reaction constant.
This equation, again, is equivalent to Eq. (\ref{D06}).




\begin{thebibliography}{33}

\expandafter\ifx\csname natexlab\endcsname\relax\def\natexlab#1{#1}\fi
\expandafter\ifx\csname bibnamefont\endcsname\relax
  \def\bibnamefont#1{#1}\fi
\expandafter\ifx\csname bibfnamefont\endcsname\relax
  \def\bibfnamefont#1{#1}\fi
\expandafter\ifx\csname citenamefont\endcsname\relax
  \def\citenamefont#1{#1}\fi
\expandafter\ifx\csname url\endcsname\relax
  \def\url#1{\texttt{#1}}\fi
\expandafter\ifx\csname urlprefix\endcsname\relax\def\urlprefix{URL }\fi
\providecommand{\bibinfo}[2]{#2}
\providecommand{\eprint}[2][]{\url{#2}}

\bibitem[{\citenamefont{Greenspan}(1978)}]{30greenspan}
\bibinfo{author}{\bibfnamefont{H.~P.} \bibnamefont{Greenspan}},
  \bibinfo{journal}{J. Fluid Mech.} \textbf{\bibinfo{volume}{84}},
  \bibinfo{pages}{125} (\bibinfo{year}{1978}).

\bibitem[{\citenamefont{Chaudhury and Whitesides}(1992)}]{15chaudhury}
\bibinfo{author}{\bibfnamefont{M.~K.} \bibnamefont{Chaudhury}}
  \bibnamefont{and} \bibinfo{author}{\bibfnamefont{G.~M.}
  \bibnamefont{Whitesides}}, \bibinfo{journal}{Science}
  \textbf{\bibinfo{volume}{256}}, \bibinfo{pages}{1539} (\bibinfo{year}{1992}).

\bibitem[{\citenamefont{Bain et~al.}(1994)\citenamefont{Bain, Burnett-Hall, and
  Montgomerie}}]{21bain}
\bibinfo{author}{\bibfnamefont{C.~D.} \bibnamefont{Bain}},
  \bibinfo{author}{\bibfnamefont{G.~D.} \bibnamefont{Burnett-Hall}},
  \bibnamefont{and} \bibinfo{author}{\bibfnamefont{R.~R.}
  \bibnamefont{Montgomerie}}, \bibinfo{journal}{Nature}
  \textbf{\bibinfo{volume}{372}}, \bibinfo{pages}{414} (\bibinfo{year}{1994}).

\bibitem[{\citenamefont{Ichimura et~al.}(2000)\citenamefont{Ichimura, Oh, and
  Nakagawa}}]{23ichimura}
\bibinfo{author}{\bibfnamefont{K.}~\bibnamefont{Ichimura}},
  \bibinfo{author}{\bibfnamefont{S.~K.} \bibnamefont{Oh}}, \bibnamefont{and}
  \bibinfo{author}{\bibfnamefont{M.}~\bibnamefont{Nakagawa}},
  \bibinfo{journal}{Science} \textbf{\bibinfo{volume}{288}},
  \bibinfo{pages}{1624} (\bibinfo{year}{2000}).

\bibitem[{\citenamefont{Bain and Whitesides}(1989)}]{19bain}
\bibinfo{author}{\bibfnamefont{C.~D.} \bibnamefont{Bain}} \bibnamefont{and}
  \bibinfo{author}{\bibfnamefont{G.~M.} \bibnamefont{Whitesides}},
  \bibinfo{journal}{Langmuir} \textbf{\bibinfo{volume}{5}},
  \bibinfo{pages}{1370} (\bibinfo{year}{1989}).

\bibitem[{\citenamefont{Schmid et~al.}(2000)\citenamefont{Schmid, Bartelt, and
  Hwang}}]{34schmid}
\bibinfo{author}{\bibfnamefont{A.~K.} \bibnamefont{Schmid}},
  \bibinfo{author}{\bibfnamefont{N.~C.} \bibnamefont{Bartelt}},
  \bibnamefont{and} \bibinfo{author}{\bibfnamefont{R.~Q.} \bibnamefont{Hwang}},
  \bibinfo{journal}{Science} \textbf{\bibinfo{volume}{290}},
  \bibinfo{pages}{1561} (\bibinfo{year}{2000}).

\bibitem[{\citenamefont{{Domingues~Dos~Santos} and
  Ondar\c{c}uhu}(1995)}]{08dossantos}
\bibinfo{author}{\bibfnamefont{F.}~\bibnamefont{{Domingues~Dos~Santos}}}
  \bibnamefont{and}
  \bibinfo{author}{\bibfnamefont{T.}~\bibnamefont{Ondar\c{c}uhu}},
  \bibinfo{journal}{Phys. Rev. Lett.} \textbf{\bibinfo{volume}{75}},
  \bibinfo{pages}{2972} (\bibinfo{year}{1995}).

\bibitem[{\citenamefont{Bico and Qu\'er\'e}(2000)}]{22bico}
\bibinfo{author}{\bibfnamefont{J.}~\bibnamefont{Bico}} \bibnamefont{and}
  \bibinfo{author}{\bibfnamefont{D.}~\bibnamefont{Qu\'er\'e}},
  \bibinfo{journal}{Europhys. Lett.} \textbf{\bibinfo{volume}{51}},
  \bibinfo{pages}{546} (\bibinfo{year}{2000}).

\bibitem[{\citenamefont{Lee et~al.}(2002)\citenamefont{Lee, Kwok, and
  Laibinis}}]{20lee}
\bibinfo{author}{\bibfnamefont{S.~W.} \bibnamefont{Lee}},
  \bibinfo{author}{\bibfnamefont{D.~Y.} \bibnamefont{Kwok}}, \bibnamefont{and}
  \bibinfo{author}{\bibfnamefont{P.~E.} \bibnamefont{Laibinis}},
  \bibinfo{journal}{Phys. Rev. E} \textbf{\bibinfo{volume}{65}},
  \bibinfo{pages}{051602} (\bibinfo{year}{2002}).

\bibitem[{\citenamefont{Wasan et~al.}(2001)\citenamefont{Wasan, Nikolov, and
  Brenner}}]{28wasan}
\bibinfo{author}{\bibfnamefont{D.~R.} \bibnamefont{Wasan}},
  \bibinfo{author}{\bibfnamefont{A.~D.} \bibnamefont{Nikolov}},
  \bibnamefont{and} \bibinfo{author}{\bibfnamefont{H.}~\bibnamefont{Brenner}},
  \bibinfo{journal}{Science} \textbf{\bibinfo{volume}{291}},
  \bibinfo{pages}{605} (\bibinfo{year}{2001}).

\bibitem[{\citenamefont{Daniel et~al.}(2001)\citenamefont{Daniel, Chaudhury,
  and Chen}}]{17daniel}
\bibinfo{author}{\bibfnamefont{S.}~\bibnamefont{Daniel}},
  \bibinfo{author}{\bibfnamefont{M.~K.} \bibnamefont{Chaudhury}},
  \bibnamefont{and} \bibinfo{author}{\bibfnamefont{J.~C.} \bibnamefont{Chen}},
  \bibinfo{journal}{Science} \textbf{\bibinfo{volume}{291}},
  \bibinfo{pages}{633} (\bibinfo{year}{2001}).

\bibitem[{\citenamefont{Bain}(2001)}]{32bain}
\bibinfo{author}{\bibfnamefont{C.~D.} \bibnamefont{Bain}},
  \bibinfo{journal}{ChemPhysChem} \textbf{\bibinfo{volume}{2}},
  \bibinfo{pages}{580} (\bibinfo{year}{2001}).

\bibitem[{\citenamefont{Kitahata and Yoshikawa}(2005)}]{03kitahta}
\bibinfo{author}{\bibfnamefont{H.}~\bibnamefont{Kitahata}} \bibnamefont{and}
  \bibinfo{author}{\bibfnamefont{K.}~\bibnamefont{Yoshikawa}},
  \bibinfo{journal}{Physica D} \textbf{\bibinfo{volume}{205}},
  \bibinfo{pages}{283} (\bibinfo{year}{2005}).

\bibitem[{\citenamefont{de~Gennes}(1998)}]{06degennes}
\bibinfo{author}{\bibfnamefont{P.~G.} \bibnamefont{de~Gennes}},
  \bibinfo{journal}{Physica A} \textbf{\bibinfo{volume}{249}},
  \bibinfo{pages}{196} (\bibinfo{year}{1998}).

\bibitem[{\citenamefont{Brochard}(1989)}]{07brochard}
\bibinfo{author}{\bibfnamefont{F.}~\bibnamefont{Brochard}},
  \bibinfo{journal}{Langmuir} \textbf{\bibinfo{volume}{5}},
  \bibinfo{pages}{432} (\bibinfo{year}{1989}).

\bibitem[{\citenamefont{Thiele et~al.}(2004)\citenamefont{Thiele, John, and
  B{\"{a}}r}}]{10thiele}
\bibinfo{author}{\bibfnamefont{U.}~\bibnamefont{Thiele}},
  \bibinfo{author}{\bibfnamefont{K.}~\bibnamefont{John}}, \bibnamefont{and}
  \bibinfo{author}{\bibfnamefont{M.}~\bibnamefont{B{\"{a}}r}},
  \bibinfo{journal}{Phys. Rev. Lett.} \textbf{\bibinfo{volume}{93}},
  \bibinfo{pages}{027802} (\bibinfo{year}{2004}).

\bibitem[{\citenamefont{Dupeyrat and Nakache}(1978)}]{11dupeyrat}
\bibinfo{author}{\bibfnamefont{M.}~\bibnamefont{Dupeyrat}} \bibnamefont{and}
  \bibinfo{author}{\bibfnamefont{E.}~\bibnamefont{Nakache}},
  \bibinfo{journal}{Bioelectrochem. Bioenerg.} \textbf{\bibinfo{volume}{5}},
  \bibinfo{pages}{134} (\bibinfo{year}{1978}).

\bibitem[{\citenamefont{Nakache et~al.}(1983)\citenamefont{Nakache, Dupeyrat,
  and Vignes-Adler}}]{33nakache}
\bibinfo{author}{\bibfnamefont{E.}~\bibnamefont{Nakache}},
  \bibinfo{author}{\bibfnamefont{M.}~\bibnamefont{Dupeyrat}}, \bibnamefont{and}
  \bibinfo{author}{\bibfnamefont{M.}~\bibnamefont{Vignes-Adler}},
  \bibinfo{journal}{J. Colloid Interface Sci.} \textbf{\bibinfo{volume}{94}},
  \bibinfo{pages}{187} (\bibinfo{year}{1983}).

\bibitem[{\citenamefont{Kai et~al.}(1985)\citenamefont{Kai, Ooichi, and
  Imaseki}}]{14kai}
\bibinfo{author}{\bibfnamefont{S.}~\bibnamefont{Kai}},
  \bibinfo{author}{\bibfnamefont{E.}~\bibnamefont{Ooichi}}, \bibnamefont{and}
  \bibinfo{author}{\bibfnamefont{M.}~\bibnamefont{Imaseki}},
  \bibinfo{journal}{J. Phys. Soc. Jpn.} \textbf{\bibinfo{volume}{54}},
  \bibinfo{pages}{1274} (\bibinfo{year}{1985}).

\bibitem[{\citenamefont{Kai et~al.}(1991)\citenamefont{Kai, M{\"{u}}ller, Mori,
  and Miki}}]{13kai}
\bibinfo{author}{\bibfnamefont{S.}~\bibnamefont{Kai}},
  \bibinfo{author}{\bibfnamefont{S.~C.} \bibnamefont{M{\"{u}}ller}},
  \bibinfo{author}{\bibfnamefont{T.}~\bibnamefont{Mori}}, \bibnamefont{and}
  \bibinfo{author}{\bibfnamefont{M.}~\bibnamefont{Miki}},
  \bibinfo{journal}{Physica D} \textbf{\bibinfo{volume}{50}},
  \bibinfo{pages}{412} (\bibinfo{year}{1991}).

\bibitem[{\citenamefont{Yoshikawa and Magome}(1993)}]{01yoshikawa}
\bibinfo{author}{\bibfnamefont{K.}~\bibnamefont{Yoshikawa}} \bibnamefont{and}
  \bibinfo{author}{\bibfnamefont{N.}~\bibnamefont{Magome}},
  \bibinfo{journal}{Bull. Chem. Soc. Jpn.} \textbf{\bibinfo{volume}{66}},
  \bibinfo{pages}{3352} (\bibinfo{year}{1993}).

\bibitem[{\citenamefont{Magome and Yoshikawa}(1996)}]{02magome}
\bibinfo{author}{\bibfnamefont{N.}~\bibnamefont{Magome}} \bibnamefont{and}
  \bibinfo{author}{\bibfnamefont{K.}~\bibnamefont{Yoshikawa}},
  \bibinfo{journal}{J. Phys. Chem.} \textbf{\bibinfo{volume}{100}},
  \bibinfo{pages}{19102} (\bibinfo{year}{1996}).

\bibitem[{\citenamefont{Nakata et~al.}(1998)\citenamefont{Nakata, Takemura, and
  Hayashima}}]{05nakata}
\bibinfo{author}{\bibfnamefont{S.}~\bibnamefont{Nakata}},
  \bibinfo{author}{\bibfnamefont{K.}~\bibnamefont{Takemura}}, \bibnamefont{and}
  \bibinfo{author}{\bibfnamefont{Y.}~\bibnamefont{Hayashima}},
  \bibinfo{journal}{Forma} \textbf{\bibinfo{volume}{13}}, \bibinfo{pages}{387}
  (\bibinfo{year}{1998}).

\bibitem[{\citenamefont{Yoshikawa and Matsubara}(1984)}]{yoshikaw1984}
\bibinfo{author}{\bibfnamefont{K.}~\bibnamefont{Yoshikawa}} \bibnamefont{and}
  \bibinfo{author}{\bibfnamefont{Y.}~\bibnamefont{Matsubara}},
  \bibinfo{journal}{J. Am. Chem. Soc.} \textbf{\bibinfo{volume}{106}},
  \bibinfo{pages}{4423} (\bibinfo{year}{1984}).

\bibitem[{\citenamefont{Shioi et~al.}(2003)\citenamefont{Shioi, Katano, and
  Onodera}}]{31shioi}
\bibinfo{author}{\bibfnamefont{A.}~\bibnamefont{Shioi}},
  \bibinfo{author}{\bibfnamefont{K.}~\bibnamefont{Katano}}, \bibnamefont{and}
  \bibinfo{author}{\bibfnamefont{Y.}~\bibnamefont{Onodera}},
  \bibinfo{journal}{J. Colloid Interface Sci.} \textbf{\bibinfo{volume}{266}},
  \bibinfo{pages}{415} (\bibinfo{year}{2003}).

\bibitem[{\citenamefont{Sumino et~al.}(2005)\citenamefont{Sumino, Magome,
  Hamada, and Yoshikawa}}]{03sumino}
\bibinfo{author}{\bibfnamefont{Y.}~\bibnamefont{Sumino}},
  \bibinfo{author}{\bibfnamefont{N.}~\bibnamefont{Magome}},
  \bibinfo{author}{\bibfnamefont{T.}~\bibnamefont{Hamada}}, \bibnamefont{and}
  \bibinfo{author}{\bibfnamefont{K.}~\bibnamefont{Yoshikawa}},
  \bibinfo{journal}{Phys. Rev. Lett.} \textbf{\bibinfo{volume}{94}},
  \bibinfo{pages}{068301} (\bibinfo{year}{2005}).

\bibitem[{\citenamefont{Mato et~al.}(1975)\citenamefont{Mato, Losada,
  Nanjundiah, and Konijn}}]{mato_1975}
\bibinfo{author}{\bibfnamefont{J.~M.} \bibnamefont{Mato}},
  \bibinfo{author}{\bibfnamefont{A.}~\bibnamefont{Losada}},
  \bibinfo{author}{\bibfnamefont{V.}~\bibnamefont{Nanjundiah}},
  \bibnamefont{and} \bibinfo{author}{\bibfnamefont{T.~M.}
  \bibnamefont{Konijn}}, \bibinfo{journal}{Proc. Nat. Acad. Sci. USA}
  \textbf{\bibinfo{volume}{72}}, \bibinfo{pages}{4991} (\bibinfo{year}{1975}).

\bibitem[{\citenamefont{Parent and Devreotes}(1999)}]{carole_1999}
\bibinfo{author}{\bibfnamefont{C.~A.} \bibnamefont{Parent}} \bibnamefont{and}
  \bibinfo{author}{\bibfnamefont{P.~N.} \bibnamefont{Devreotes}},
  \bibinfo{journal}{Science} \textbf{\bibinfo{volume}{284}},
  \bibinfo{pages}{765} (\bibinfo{year}{1999}).

\bibitem[{\citenamefont{Atkin et~al.}(2003)\citenamefont{Atkin, Craig, Wanless,
  and Biggs}}]{atkin03}
\bibinfo{author}{\bibfnamefont{R.}~\bibnamefont{Atkin}},
  \bibinfo{author}{\bibfnamefont{V.~S.~J.} \bibnamefont{Craig}},
  \bibinfo{author}{\bibfnamefont{E.~J.} \bibnamefont{Wanless}},
  \bibnamefont{and} \bibinfo{author}{\bibfnamefont{S.}~\bibnamefont{Biggs}},
  \bibinfo{journal}{Adv. Colloid Interf. Sci.} \textbf{\bibinfo{volume}{103}},
  \bibinfo{pages}{219} (\bibinfo{year}{2003}).

\bibitem[{\citenamefont{Parks}(1965)}]{01parks}
\bibinfo{author}{\bibfnamefont{G.~A.} \bibnamefont{Parks}},
  \bibinfo{journal}{Chem. Rev.} \textbf{\bibinfo{volume}{65}},
  \bibinfo{pages}{177} (\bibinfo{year}{1965}).

\bibitem[{\citenamefont{Yoon et~al.}(1979)\citenamefont{Yoon, Salmam, and
  Donnay}}]{02yoon}
\bibinfo{author}{\bibfnamefont{R.~H.} \bibnamefont{Yoon}},
  \bibinfo{author}{\bibfnamefont{T.}~\bibnamefont{Salmam}}, \bibnamefont{and}
  \bibinfo{author}{\bibfnamefont{G.}~\bibnamefont{Donnay}},
  \bibinfo{journal}{J. Colloid Interface Sci.} \textbf{\bibinfo{volume}{70}},
  \bibinfo{pages}{483} (\bibinfo{year}{1979}).

\bibitem[{\citenamefont{Shanahan and de~Gennes}(1997)}]{09shanahan}
\bibinfo{author}{\bibfnamefont{M.~E.~R.} \bibnamefont{Shanahan}}
  \bibnamefont{and} \bibinfo{author}{\bibfnamefont{P.~G.}
  \bibnamefont{de~Gennes}}, \bibinfo{journal}{C. R. Acad. Sci. Paris II}
  \textbf{\bibinfo{volume}{324}}, \bibinfo{pages}{261} (\bibinfo{year}{1997}).

\bibitem[{\citenamefont{Shanahan}(2000)}]{25shanahan}
\bibinfo{author}{\bibfnamefont{M.~E.~R.} \bibnamefont{Shanahan}},
  \bibinfo{journal}{J. Colloid Interface Sci.} \textbf{\bibinfo{volume}{229}},
  \bibinfo{pages}{168} (\bibinfo{year}{2000}).

\end{thebibliography}
\end{document}